\documentclass[apj]{emulateapj}
\usepackage{graphicx}

\def\blender{{\tt BLENDER}}
\def\kepler{\emph{Kepler}}
\def\spitzer{\emph{Spitzer}}
\def\corotseven{CoRoT-7}
\def\corot{CoRoT}
\def\modif{}

\begin{document}

\title{Spitzer Infrared observations and independent validation of the
transiting Super-Earth \corotseven\,b.}

\author{
Francois Fressin\altaffilmark{1}, 
Guillermo Torres\altaffilmark{1}, 
Frederic Pont\altaffilmark{2}, 
Heather A.\ Knutson\altaffilmark{3}, 
David Charbonneau\altaffilmark{1}, 
Tsevi Mazeh\altaffilmark{4}, 
Suzanne Aigrain\altaffilmark{5}, 
Malcolm Fridlund\altaffilmark{6}, 
Christopher E.\ Henze\altaffilmark{7},
Tristan Guillot\altaffilmark{8}, and
Heike Rauer\altaffilmark{9}
}

\altaffiltext{1}{Harvard-Smithsonian Center for Astrophysics, 60
Garden St., Cambridge, MA 02138, USA, e-mail:
ffressin@cfa.harvard.edu}
\altaffiltext{2}{University of Exeter, EX4 4QL, Exeter, UK}
\altaffiltext{3}{University of California, Berkeley, CA 94720, USA}
\altaffiltext{4}{Tel Aviv University, 69978 Tel Aviv, Israel}
\altaffiltext{5}{University of Oxford, OX1 3RH, Oxford, UK}
\altaffiltext{6}{ESTEC/ESA, PO Box 299, 2200 AG Noordwijk, The Netherlands} 
\altaffiltext{7}{NASA Ames Research Center, Moffett Field, CA 94035, USA}
\altaffiltext{8}{Observatoire de la C\^ote d'Azur, BP 4229, 06304, Nice, France}
\altaffiltext{9}{Deutsches Zentrum f\"ur Luft und Raumfahrt (DLR), Rutherfordstr.\ 2, 12489, Berlin, Germany}

\begin{abstract}

The detection and characterization of the first transiting
super-Earth, \corotseven\,b, has required an unprecedented effort in
terms of telescope time and analysis.  Although the star does display
a radial velocity signal at the period of the planet, this has been
difficult to disentangle from the intrinsic stellar variability, and
pinning down the velocity amplitude has been very challenging.  As a
result, the precise value of the mass of the planet --- and even the
extent to which it can be considered to be confirmed --- have been
debated in the recent literature, with six mass measurements
published so far based on the same spectroscopic observations, ranging
from about 2 to 8 Earth masses.

Here we report on an independent validation of the planet discovery,
using one of the fundamental properties of a transit signal: its
achromaticity. We observed four transits of \corotseven\,b at
4.5\,\micron\ and 8.0\,\micron\ with the Infrared Array Camera (IRAC)
onboard the \spitzer\ Space Telescope, in order to determine whether
the depth of the transit signal in the near-infrared is consistent with
that observed in the \corot\ bandpass, as expected for a planet. We
detected the transit and found an average depth of $0.426 \pm 0.115$
mmag at 4.5\,\micron, which is in good agreement with the depth of
$0.350 \pm 0.011$ mmag (ignoring limb darkening) found by \corot.  The
observations at 8.0\,\micron\ did not yield a significant detection.
The 4.5\,\micron\ observations place important constraints on the
kinds of astrophysical false positives that could mimic the
signal. Combining this with additional constraints reported earlier,
we performed an exhaustive exploration of possible blends scenarios
for \corotseven\,b using the \blender\ technique. We are able to rule
out the vast majority of false positives, and the remaining ones are
found to be much less likely than a true transiting planet. We thus
validate \corotseven\,b as a bona-fide planet with a very high degree
of confidence, independently of any radial-velocity information.  Our
\spitzer\ observations have additionally allowed us to significantly
improve the ephemeris of the planet, so that future transits should be
recoverable well into the next decade.

In its warm phase \spitzer\ is expected to be an essential tool for
the validation, along the lines of the present analysis, of transiting
planet candidates with shallow signals from \corot\ as well as from
the \kepler\ Mission, including potentially rocky planets in the
habitable zones of their parent stars.
                                                               
\end{abstract}


\section{Introduction}\label{sec:intro}

Among the known exoplanets, a few special cases stand out as the
objects that inaugurated the study of the physics of Earth-like
exoplanets.  \corotseven\,b \citep{lege:09} is the first super-Earth
for which the mass and radius have been estimated, and provided the
first real constraints on models of the formation, structure, and
evolution of small and potentially rocky exoplanets.  Kepler-10\,b has
recently been announced as the first rocky planet found by the
\kepler\ Mission (Batalha et al.\ 2010), with a mass of $M_p =
4.6^{+1.2}_{-1.3}\,M_{\earth}$ and a radius of $R_p =
1.416^{+0.033}_{-0.036}\,R_{\earth}$. We now also have examples of
interesting planets that are intermediate in mass and radius between
the Earth and Neptune, such as GJ\,1214\,b, with $M_p = 6.55 \pm
0.98\,M_{\earth}$ (Charbonneau et al.\ 2009) and $R_p = 2.64 \pm
0.13\,R_{\earth}$ \citep{bert:11}.

The first detection of a super-Earth was made possible by the
successful \corot\ satellite \citep{bagl:02}. The discovery was made
in the course of observations in the first long run of this mission in
the direction of the anti-center of the Galaxy (LRa01), which took
place from 2007 October to 2008 March. A small transit-like signal was
identified with a depth of 0.35~mmag and duration of 1.3~hr, recurring
with a period of 0.8535~days and being consistent with a super-Earth
size planet orbiting a bright ($V=11.7$, $K=9.8$) G9 dwarf star.  The
discovery triggered a series of follow-up observations to clarify the
origin of the shallow transit signal.  The detection and the follow-up
campaign have been fully described by \cite{lege:09}.  Bruntt et al.\
(2010) subsequently performed a detailed spectroscopic analysis of the
\corotseven\ star and determined an improved stellar radius of
$R_\star = 0.82 \pm 0.04\,R_{\sun}$. This resulted in a revised planet
radius of $R_p = 1.58 \pm 0.10\,R_{\earth}$.

\cite{quel:09} reported on an extensive observational campaign carried
out with the HARPS instrument on the ESO 3.6\,m telescope at La Silla,
with the goal of detecting the Doppler signal of this small object and
measuring its mass. \corotseven\ is an active star, however, with
starspot-induced photometric variability at the $\sim$2\% percent
level modulated by the stellar rotation period of about 23 days.  The
radial velocity of the star is dominated by an irregular signal with
an amplitude several times larger than the sought-after signature of
the transiting planet. Under these circumstances, measuring the mass
accurately is a challenging task that depends strongly on how the
activity-induced variability is handled, on assumptions about possible
additional non-transiting planets that may be contributing (up to two
have been considered), and on the attitude toward systematic errors.
\cite{quel:09} produced the first mass estimate of $M_p = 4.8 \pm
0.8\,M_{\earth}$. Subsequent authors have reported different values
using \emph{the same HARPS observations} or subsets thereof. These
estimates are not always consistent with each other within their
formal errors, and vary considerably in significance level: $6.9 \pm
1.4\,M_{\earth}$ \citep{hatz:10}, $2.3 \pm 1.5\,M_{\earth}$
\citep{pont:11}, $8.0 \pm 1.2\,M_{\earth}$ \citep{Ferraz-Mello:11},
$5.7 \pm 2.5\,M_{\earth}$ \citep{Boisse:11}, and $7.4 \pm
1.2\,M_{\earth}$ \citep{hatz:11}.  The impact of the discrepancies is
not insignificant, as these estimates lead to rather different mean
densities and therefore different internal structures for the planet
as inferred from current theory.  While some of the more recent
determinations appear to favor a higher mass for the planet, at the
2$\sigma$ level the estimates range all the way from 0 to
10\,$M_{\earth}$, and for some of the lower values \citep{pont:11,
Boisse:11} the statistical significance of the Doppler detection of
\corotseven\,b is considerably less compelling.

Similar difficulties are expected to occur for other candidates with
shallow transits, in which ``confirmation'' in the usual sense of the
word by the detection of a radial-velocity signature that is at the
limit of detectability with current instrumentation will be very
challenging.  With the recent announcement of a large number of
shallow transit candidates discovered by the \kepler\ satellite
\citep{boru:11}, obtaining assurance that these signals correspond to
bona-fide super-Earth size planets, as opposed to a false positive, is
among the most urgent tasks that lie ahead, and is of primary
importance for the statistical interpretation of the results.

Motivated by the lingering problems with the \corotseven\,b mass
determination described above, and the implications for the robustness
of the detection, we use this case here to illustrate the application
of a powerful technique to help ``validate''\footnote{In the context
of this paper ``confirmation'' as used above refers to the unambiguous
detection of the gravitational influence of the planet on its host
star (e.g., the Doppler signal) to establish the planetary nature of
the candidate; when this is not possible, we speak of ``validation'',
which involves an estimate of the false alarm probability.} shallow
transit candidates \emph{independently of any radial-velocity
information}. It makes use of near-infrared observations with the
\spitzer\ Space Telescope, and is based on the premise that true
transits are achromatic signals, to first order, so that the transit
depth as observed in the near-infrared should be the same as in the
\corot\ passband (ignoring the effects of limb darkening).  If the
candidate is the result of a blend, however, the depth can be
significantly different. {\modif As an example, if the observed
transit were due to an object eclipsing a background star of
$0.5\,M_{\odot}$ and sun-like age and metallicity, then its transit in
the Spitzer 4.5\,\micron\ bandpass would be 3.2 times deeper than in
the \corot\ bandpass.} With its infrared passband, \spitzer\ affords
the maximal wavelength separation from the \corot\ photometry
(passband around 650\,nm), and places very strong constraints on
possible false positive scenarios, as we describe below.

Even with these constraints, and others available from follow-up
observations carried out and reported by the \corot\ team, it is not
possible to rule out all possible blend configurations for
\corotseven\,b, as recognized also by \cite{lege:09}. Thus, the main
goal of this work is to more exhaustively explore the wide variety of
false positive scenarios that can mimic the light curve, {\modif to
obtain a realistic estimate of the blend frequency that may be
expected.  We aim to provide an independent assessment of the
confidence level that the signal is of planetary nature. Rather than
focusing solely on the likelihood of a blend (frequentist approach),
as in previous studies, we adopt a Bayesian approach in which we
compare the blend frequency with a prior for the likelihood of a
planet (odds ratio). To evaluate the blend frequency} we make use of
the \blender\ technique introduced by \cite{Torres:04, Torres:11},
with further developments as described by \cite{Fressin:11}. This
methodology has been applied successfully to validate a number of
shallow transit signals from the \kepler\ Mission {\modif including
Kepler-9\,d, Kepler-10\,c, Kepler-11\,g, and Kepler-19\,b
\citep{Torres:11, Fressin:11, Lissauer:11, Ballard:11}.}

We begin by describing our \spitzer\ observations {\modif
(Section~\ref{sec:spitzer})}, and then briefly summarizing the data
used here along with other follow-up observations relevant to this
investigation {\modif (Section~\ref{sec:photometry})}. This is
followed by the \blender\ analysis that examines the vast space of
parameters for false positives by synthesizing realistic blend light
curves and comparing them with the \corot\ photometry {\modif
(Section~\ref{sec:blender})}.  We next estimate the expected frequency of
blends and compare it with the expected frequency of planets {\modif
(Section~\ref{sec:validation})}. As shown below, this analysis is able
to validate \corotseven\,b as a planet without relying on any
radial-velocity information.

\section{\spitzer\ observations of \corotseven\,b}\label{sec:spitzer}

The Infrared Array Camera \citep[IRAC;][]{fazi:04} of the \spitzer\
Space Telescope \citep{wern:04} obtains simultaneous images in four
bandpasses. A $5\farcm2 \times 5\farcm2$~field of view (FOV) is imaged
in one pair of bandpasses (3.6 and 5.8\,\micron), and a nearly
adjacent FOV imaged in the second pair (4.5 and 8.0\,\micron). The two
blue channels employ InSb detectors, whereas the red channels use
Si:As IBC detectors.  All four arrays are $256\times256$ pixels.
{\modif While the present Warm \spitzer\ mission is restricted to the
two shorter wavelengths, the data discussed in this section were
obtained just prior to the spacecraft entering that phase in May of
2009.}  We elected to monitor \corotseven\ in only one channel pair
(4.5 and 8.0\,\micron), as even if the stellar flux in these
bandpasses is slightly smaller, we have obtained more precise
observations in the past using this pair \citep{knut:08, fres:09}. It
is also farther in wavelength from the \corot\ bandpass, and the
expected depth difference if the signal comes from a blend scales with
the wavelength difference{\modif, as we describe later}.

We used IRAC to observe the primary eclipse of \corotseven\,b on UT
2009 April 22, 23, 24 and 25, obtaining data at 4.5 and
8.0\,\micron. We were able to observe it in full array mode in both
channels for a total duration of 17.0 hours, {\modif including four
transits lasting 75 minutes each}.  We observed the target in the IRAC
stellar mode, in which the camera gathers simultaneously two 10.4\,s
integrations at 4.5 and 8.0\,\micron. Therefore, we gathered
respectively 1075, 1212, 1212, and 1212 images both at 4.5 and
8.0~\micron\ during the four transits we observed. Our goal was to
detect the transit signal at 4.5\,\micron, but as the 8.0\,\micron\
observations are simultaneous and automatic, we present them hereafter
for completeness. We describe below our observations in two sections,
as the InSb detectors used for IRAC channels at 4.5\,\micron\ require
a different treatment than the Si:As detectors of the IRAC
8.0\,\micron\ channel.

We were mindful that the signal we were looking for was at the limit
of what was possible to obtain with a few transits from \spitzer, as
it has mainly been used to look at brighter stars and deeper eclipses.
Our expectation in terms of statistical significance for the detection
of a single transit was $2.1\sigma$ at 4.5\,\micron\ ($4.2\sigma$
when combining the four transits), scaled on the detection level we
achieved in our previous studies of TrES-1, TrES-3, and TrES-4
\citep{char:08,knut:08,fres:09}.

\subsection{4.5\,\micron\ observations}\label{sec:short_norm}

The contribution of the background to the total flux from \corotseven\
is low in the 4.5~\micron\ IRAC bandpass, contributing only 0.35\% to
the total flux in an aperture with a 5-pixel diameter centered on the
position of the star. We obtained the lowest RMS time series using an
aperture with a radius of 5.0 pixels. We allowed the position of our
aperture to shift with the position of the star in each image.  We
estimated the background in each image from an annulus with an inner
radius of 12 pixels and an outer radius of 20 pixels centered on the
position of the star.

We determined the position of the star in each image by fitting a
two-dimensional Gaussian to the position of the star. Agol et
al.\ (2010) have recently completed a comparative analysis of different
methods to estimate the stellar centroid in \spitzer\ images. Their
best results are obtained by fitting a two-dimensional Gaussian to the
star PSF, compared to a flux-weighted centroiding, and parabolic
fitting.  This Gaussian algorithm uses the $7\times7$ pixel sub-array
from the image centered on the brightest pixel. It fits a
two-dimensional Gaussian to this array, allowing its center,
amplitude, and width to vary. It then uses a non-linear
Levenberg-Marquardt algorithm to optimize these parameters
\citep{mark:09}. We compared this technique with the position-weighted
sum of the flux in a 4-pixel radius disk centered on the approximate
position of the star, which we previously used on similar magnitude
targets \citep{char:08,knut:08,fres:09}. The Gaussian fit proved to be
better regarding two criteria. First, the scatter between the position
estimates was 1.3 times smaller. Second, the scatter of the
differential position between our target star and a nearby reference
star (2 magnitudes fainter in channel 2 and at 28\arcsec) was smaller
and did not show any correlation with the intra-pixel position.

The dominant instrumental systematic effect in the first two IRAC
bandpasses is due to a well-known intra-pixel sensitivity
\citep{reac:05,char:05,char:08,mor:06,knut:08}. Fluxes at these two
wavelengths show a strong correlation with the intra-pixel position of
the star on the detector, at a level comparable to the expected depth
of the eclipse. We used the following parameters to fit the observed
flux as a linear function of the subpixel position:
\begin{equation}\label{eq1}
f=f_0*\left[c_1+c_2(x-x_0)+c_3(y-y_0)\right]\,,
\end{equation}
where $f_0$ is the original flux from the star, $f$ is the measured
flux, $x$ and $y$ denote the location of the Gaussian-fit centroid of
the star on the array, $x_0$ and $y_0$ are the coordinates of the
center of the pixel containing the peak of the star's point spread
function, and $c_1$--$c_3$ are free parameters in the fit.  {\modif We
excluded the in-transit measurements, based on the known ephemeris, in
order to avoid suppressing the transit depth.}  {\modif For the $x$
and $y$ positions above} we calculated the centroid of the target in
each image and found that the pointing jitter was around 0.12 pixels
($0\farcm14$) over the course of a visit in both $x$ and $y$. The
pointing drift of the telescope appears to occur on a longer timescale
than the exposures.  Therefore, instead of using the actual position
estimate at individual exposure times, we smoothed the $x$ and $y$
curves as a function of time and used the smoothed position instead.

In contrast to previous observations of HD\,189733 and HD\,209458
\citep{knut:08, char:08}, we found that adding quadratic or
higher-order terms to this equation, or even cross-terms, did not
improve the fit significantly, likely due to the lower signal-to-noise
ratio of the present observations.

After correcting for the intra-pixel sensitivity, a decreasing trend
was still visible that is likely to be an instrumental effect related
to the detector or telescope, and has also been seen in observations
of TrES-3 and TrES-4 \citep{knut:08, fres:09}, two stars with similar
brightness. We corrected for this effect by fitting the data in both
channels with a linear function of time.  This term was fitted
simultaneously with the correction for the intra-pixel sensitivity, so
that we can accurately characterize the additional uncertainty in the
depth and timing of the eclipse introduced by these corrections. That
is to say, we solved for four parameters including a constant term, a
linear function of $x$ position, a linear function of $y$ position,
and a linear function of time. We also trimmed the first 30 minutes of
data that show a larger scatter, as we have done in previous cases.
The fit was performed with a Markov Chain Monte Carlo (MCMC) method
\citep{ford:05,winn:07} with $5 \times 10^5$ steps, where we set the
uncertainty of the individual measurements equal to the standard
deviation of the out-of-transit data after correction for the various
detector effects.

Prior to the fit we carried out an initial trimming within our
aperture, discarding outliers farther than 3.5$\sigma$ from the local
median flux (defined as the median of a 15-minute window centered on
the data point). We also removed measurements for which the identified
position of the photocenter $x$ or $y$ deviated by more than
3.5$\sigma$ from the same 15-minute median position. This global
trimming excludes 6.6\% of the data points in the four visits in the
4.5\,\micron\ bandpass.  We excluded outliers greater than 3.5$\sigma$
during each step of the chain, as determined using the residuals from
the model light curve, from our evaluation of the $\chi^2$
function. We rescaled the value of the $\chi^2$ function to account
for the fact that we are varying the number of pixels included in the
fit.

\begin{figure}
\epsscale{0.65}
\plotone{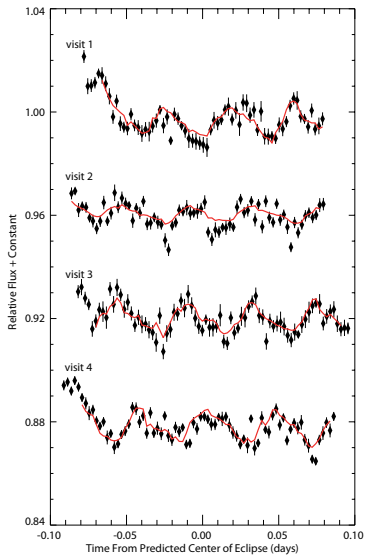}
\caption{Transit light curves of \corotseven\,b in channel~2
  (4.5\,\micron) observed on UT 2009 April 22, 23, 24 and 25, with
  best-fit curves representing instrumental effects overplotted in
  red. Data have been binned in 7.3 minute intervals, then offset by a
  constant for the purposes of this plot.\label{four_eclipses}}
\end{figure}

After producing the chain, we searched for the point in the chain
where the $\chi^2$ value first falls below the median of all the
$\chi^2$ values (i.e., where the code had first found the best-fit
solution), and discarded all the steps up to that point.
Figure~\ref{four_eclipses} shows the four individual light curves and
the respective fits for the instrumental effects that we removed from
these curves before looking for the transit signal itself.

Next we carried out a second Markov chain fit of the transit signal
itself on the trimmed data. We initially allowed the individual depths
and times of the four transits to vary independently, along with the
normalized semimajor axis $a/R_\star$, and the inclination angle of
the orbit.  We calculated our transit curves using the formulation by
\citet{mand:02}. Although small at 4.5\,\micron, limb darkening was
taken into account in our modeling according to the four-parameter law
by \cite{Claret:00}, with coefficients taken from the work of
\cite{Sing:10} {\modif and stellar parameters $T_{\rm eff} = 5250 \pm
60$\,K, $\log g = 4.47 \pm 0.05$, and ${\rm [Fe/H]} = +0.12 \pm 0.06$
from \cite{brun:10}}.  While the individual depths and transit time
parameters were consistent with transits occurring at the expected
period from the \corot\ light curve, the signal-to-noise ratios were
too poor to provide a meaningful constraint on the geometric
parameters.  We therefore chose to restrict the space of free
parameters for the Markov chain analysis to the transit depth,
$a/R_\star$, inclination angle, and period, and we held the transit
epoch fixed at the value reported by \cite{lege:09}.

Non-random sources of noise in transit and eclipse photometry --- such
as instrumental systematics and stellar variability --- could dominate
the error budget in the derived system parameters. This is true for
ground-based data, and also turns out to be true for space-based
data. The higher stability of space measurements is offset by the fact
that smaller effects are being measured, and correspondingly smaller
levels of random error are achieved by collecting more signal.
\cite{pont:06} showed how neglecting this type of noise could lead to
an underestimate of the actual uncertainties by a large factor. In the
present study, a realistic error estimate is important and we
therefore attempted to assess the possible impact of non-random
noise. The presence of correlated noise is obvious in our raw
photometric sequence. We corrected for it to first order, but
obviously the correction cannot be perfect. We used the
single-parameter description of the correlated component of the noise
proposed by \cite{pont:06}, with the further simplification of
\cite{winn:09} adapted to regularly-sampled data, to estimate the
impact of the residual systematics after decorrelation.  We repeated
the MCMC analysis by using modified uncertainties $\sigma_{\rm tot}^2
= \sigma_{\rm w}^2 + n \sigma_{\rm r}^2$, where $\sigma_{\rm w}^2$ and
$\sigma_{\rm r}^2$ are the `white' and `red' components of the noise
(i.e., random and correlated, respectively), and $n$ is the number of
data points during the eclipse. We estimated $\sigma_{\rm r}$ from the
dispersion of the flux after decorrelation obtained with slightly
different, reasonable decorrelation procedures.  We used
$\sigma_{r}=0.000108$\,mag as the dispersion between three different
decorrelation techniques (i.e., fitting the intra-pixel answer and the
transit simultaneously, adding quadratic terms to the intra-pixel
answer, and the decorrelation technique previously described).

\begin{figure}
\epsscale{0.9}
\plotone{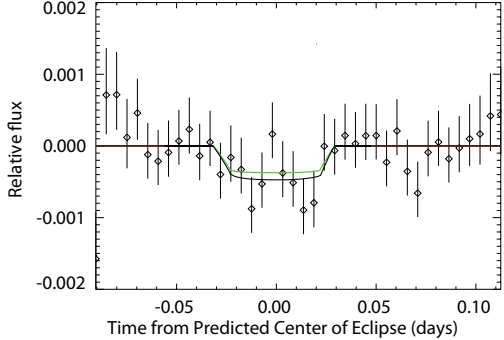}

\caption{Transit light curve of \corotseven\,b in the IRAC
  4.5\,\micron\ bandpass after removal of instrumental effects, folded
  on the expected period. Data have been normalized to remove detector
  effects (see discussion in Sect.~\ref{sec:short_norm}), and binned
  in 9.2 minute intervals. The black curve is the best transit model
  fit to the \spitzer\ light curve, imposing the \corot\
  ephemeris. {\modif The transit depth is $0.426 \pm 0.115$\,mmag.}
  The green curve shows the model expected for the super-Earth transit
  scenario, adopting the \corot\ parameters from \cite{lege:09},
  {\modif which lead to a depth of $0.350 \pm 0.011$\,mmag}. The good
  agreement in the depth indicates the transit is achromatic, as
  expected for a planet.\label{folded}}
\end{figure}

Table \ref{eclipse_depths} collects our results, and the \spitzer\
time series is shown in Figure~\ref{folded} along with the fitted
model.  The parameters we derive for the planet are in good agreement
with those based on the light curve as observed by the \corot\
satellite.  In particular, the near-infrared and optical depths are
consistent within the errors. Also shown in Figure~\ref{folded} is a
model based on the optical lightcurve parameters, for comparison with
the \spitzer\ curve.  Although we do not reach the precision
\cite{lege:09} obtained for the geometric parameters based on their
detection of 153 individual transits, we were able to improve the
precision in the period determination of the planet significantly, as
our \spitzer\ observations were gathered some 640 planetary orbits
after the original \corot\ data.

\begin{deluxetable}{lcc}
\tablecaption{Best-fit transit parameters \label{eclipse_depths}}
\tablewidth{0pt}
\tablehead{
\colhead{Parameter} & \colhead{\corot\tablenotemark{a}}  & \colhead{\spitzer\ 4.5\,\micron}}
\startdata
$P$ (day) & $0.853585\pm0.000024$ & $0.853590\pm0.000006$\\
$a/R_\star$ & $4.27\pm0.20$ & $4.1_{-1.6}^{+2.4}$\\
Inclination (deg) & $80.1\pm0.3$ & $83.6_{-8.3}^{+6.4}$\\
Depth (mmag) & $0.350\pm0.011$ & $0.426\pm0.115$ \\ 
\enddata
\tablenotetext{a}{\cite{lege:09}.}
\end{deluxetable}

To put our results on a more quantitative basis, we computed the
reduced $\chi^2$ values corresponding to the case where no transit at
all is shown in the data, and the case of a transit with a depth
corresponding to a planet such as \corotseven\,b. We find that the
super-Earth scenario gives a reduced $\chi^2$ (1.0039) very close to
the best fit to the data (1.0033), whereas the no-transit scenario
provides a poor fit (1.0276).

Since the transit depth we measure in the near infrared is about the
same as in the optical, this argues against blends composed of stars
of much later spectral type. Importantly, we can use the error bar on
the measured depth to rule out blends involving stars of a different
temperature that would have produced a significantly different depth
in infrared.  {\modif Indeed, the depth of a blend in the \spitzer\
bandpass compared to the optical depends solely on the infrared
contribution of the blending star to the total flux relative to its
contribution in the \corot\ bandpass. For example, if the relative
contribution in the infrared of a background eclipsing binary is three
times larger in infrared than in the visible, then the observed
transit in the infrared would be three times deeper.
The relative infrared-to-visible contribution of a background star is
also independent of its distance to the target, to first order (i.e.,
ignoring the effects of interstellar dust). As a consequence, for a
given target star, the observed depth increase or decrease of a blend
when observed in the \spitzer\ bandpass is only a function of the
spectral type of the contaminating star. Using an isochrone of solar
age and metallicity representative of the background stars in the
Galactic plane, it is straightforward to compute the change in the
relative infrared-to-visible flux contribution of such a star as a
function of its mass, and hence the depth increase or decrease in the
4.5 and 8\,$\micron$ bandpasses that a blend would display. With the
target properties (mass $M = 0.91\,M_{\sun}$, age $\approx 2$\,Gyr)
from \cite{brun:10}, we find for \corotseven\ that all blended stars
below $0.69\,M_{\sun}$ would show a transit deeper than the 3-$\sigma$
upper limit observed with \spitzer\ at 4.5\,$\micron$ (0.77\,mmag).
We make use of this important constraint later in
Sect.~\ref{sec:validation} to eliminate a large fraction of potential
false positives for \corotseven\,b.}

\subsection{8.0\,\micron\ observations}\label{sec:long_norm}

For the observations at 8.0\,\micron\ we used the ``preflash''
technique (Knutson et al.\ 2009), in which we pointed the telescope for
30 minutes towards a bright \ion{H}{2} region before observing
\corotseven. This was completed in order to reduce the amplitude of
the detector ``ramp'' at 8.0\,\micron, effectively pre-loading the
pixels on which the target star would be pointed.

Previous secondary eclipse studies (e.g., Knutson et al.\ 2008) have
shown that PSF-fitting can provide a better signal-to-noise ratio at
longer wavelengths. At longer wavelengths the flux from the star is
smaller and the zodiacal background is larger; we find that the
background contributes 20\% of the total flux in a 3-pixel aperture at
8.0\,\micron.

At 8.0\,\micron, we found that the relative scatter in the time series
after model fitting from the PSF fits was 15\% higher than in the time
series from aperture photometry with a 3.0 pixel radius. As a result
of this increased scatter, which is likely produced by discrepancies
between the model PSF and the observed PSF, we concluded that aperture
photometry is also preferable. We compared the time series using
apertures ranging from 3.0 to 4.5 pixels and found consistent results
in all cases, but with a scatter that increases with the radius of the
photometric aperture.

Previous observations \citep[e.g.,][]{knut:09} have shown that the
ramp is well described as following an asymptotic shape, with a
steeper rise in the first 30 minutes of observations. We corrected for
this effect by fitting our time series with the following function:
\begin{equation}\label{eq2}
f=f_0*\left[c_1+c_2 \ln(dt)\right]\,,
\end{equation}
where $f_0$ is the original flux from the star, $f$ is the measured
flux, and $dt$ is the elapsed time in days since the start of the
observations.

As the expected statistical significance of the signal at
8.0\,\micron\ was very small (i.e., 1.1\,$\sigma$), we
chose to fix the transit time and period of the transit to values
found in our 4.5\,\micron\ study and verified that the residual signal
in the four visits at 8.0\,\micron\ was compatible with the transiting
planet scenario.  We carried out a Markov Chain Monte Carlo fit to the
data as described in Sect.~\ref{sec:short_norm}, simultaneously
fitting Eq.~\ref{eq2} and a transit model, with the depth as the only
free parameter in the latter. No significant correlations were found
between the variables.  As a further check we repeated these fits
adding a quadratic term of the form $\ln(dt)$ in Eq.~\ref{eq2}, and
found that the value of $\chi^2$ for our best-fit solution was similar
to the previous value, so we chose not to include this additional
term.

\begin{figure}
\epsscale{0.9}
\plotone{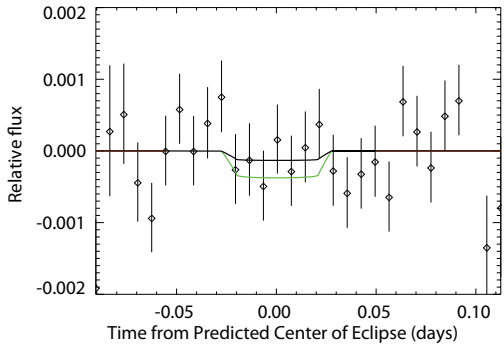}
\caption{Transit light curve of \corotseven\,b in the IRAC
  8.0\,\micron\ bandpass after removal of instrumental effects, phased
  with the expected period and with the best-fit transit curve
  overplotted (see text).  Data have been normalized to remove
  detector effects (see discussion in Sect.~\ref{sec:short_norm} and
  Sect.~\ref{sec:long_norm}), and binned in 9.2 minute intervals. The
  black curve represents the best transit model fit adopting the
  \corot\ ephemeris, {\modif which gives a statistically insignificant
    depth of $0.11 \pm 0.30$\,mmag}. The green curve shows the fit
  expected for the super-Earth transit scenario, using the parameters
  from \cite{lege:09}.
\label{folded8}}
\end{figure}

Figure~\ref{folded8} shows the detrended (from the ramp), trimmed,
binned, and folded channel~4 light curve. We estimated the best-fit
transit depth using a Markov Chain Monte Carlo fit in the same way as
we did for the 4.5\,\micron\ data, and found an eclipse depth of $0.11
\pm 0.30$\,mmag. The noise level is too large to say anything about a
dip corresponding to the planet eclipse, but it disallows a blend
scenario involving a significantly redder star. {\modif Specifically,
proceeding in the same way as described at the end of
Sect.~\ref{sec:short_norm}, a star of 0.5\,$M_{\sun}$ would be
excluded at the 3$\sigma$ level.}

\section{\corot\ photometry and follow-up observations}
\label{sec:photometry}

The photometric observations used in the next section to investigate
blends are essentially the same as used by \cite{lege:09}, with a
somewhat different de-trending of the data.  We started with the N2
(science grade) light curve, discarding the initial few days, which
had a time sampling of 512s, and worked only with the remainder, where
the time-sampling was 32s \citep[see][for a detailed description of
the N2 light curve]{lege:09}. We identified and clipped outliers using
a running 5-point median filter, and then modeled the intrinsic
stellar variability using an iterative non-linear filter
\citep{aigr:04}. This consists in applying a 5-point boxcar filter
followed by a median filter with a width of 1 day, $3 \sigma$ clipping
the residuals, and iterating until no more points are clipped. The
resulting slowly-varying component was then subtracted from the
original light curve, to give the time-series used in this work. We
made use of the white-light data only. Of the extensive follow-up
observations carried out by the \corot\ team as described by
\cite{lege:09}, the most relevant for our study are the
high-resolution imaging observations obtained with the NACO instrument
on the VLT. These data exclude any blends capable of mimicking the
signal outside of 0\farcs4 from the target down to 6.5~mag fainter
than the target. Additionally, the spectroscopic observations reported
by these authors allowed them to rule out companions that would be
bright enough to be visible in the spectrum (as a second set of
lines). In particular, their near-infrared ($K$-band) spectroscopy
with CRIRES on the VLT rules out most companions brighter than about
7\% of the flux of the target in that band ($\Delta m \approx 2.9$).
For the analysis below we adopt a more conservative limit of $\Delta m
= 1.0$.

\section{\blender\ analysis}\label{sec:blender}

Blend scenarios satisfying the CoRoT observations were explored with
the \blender\ technique \citep{Torres:04, Torres:11, Fressin:11}, by
examining the quality of the fit to the \corotseven\ photometry of a
large array of synthetic model light curves. These synthetic light
curves result from the combined light of three objects composing the
blend: the main target, and an unresolved eclipsing pair along the
line of sight with the same period as detected in \corotseven, whose
eclipses are attenuated by the light from the target. The eclipsing
system may be physically associated (hierarchical triple), or may be
in the background or foreground, and the eclipsing pair may consist of
two stars, or a star transited by a larger planet.  \blender\ uses the
detailed shape of the transit light curve to weed out scenarios that
lead to the wrong shape for a transit. The properties of the three
objects were taken from model isochrones by \cite{gira:00} in order to
synthesize light curves, and {\modif these artificial light curves}
were compared with the observations in a $\chi^2$ sense {\modif (i.e.,
by computing the sum of the squared residuals normalized by the
photometric uncertainties)}.  For practical reasons blend light curves
were calculated here in the \kepler\ passband, which is sufficiently
similar to the passband of the \corot\ satellite for our purposes, in
both the central wavelength and width.\footnote{(see {\tt
http://keplergo.arc.nasa.gov/CalibrationResponse.shtml})} The
properties of the main star {\modif (essentially its intrinsic
brightness, which affects the dilution of the eclipses)} were
constrained by the spectroscopic observations of \corotseven\
\citep[{\modif $T_{\rm eff} = 5250 \pm 60$\,K, $\log g = 4.47 \pm
0.05$, ${\rm [Fe/H]} = +0.12 \pm 0.06$};][]{quel:09, brun:10}, and
held fixed. Those of the other two objects (referred to here as the
`secondary' {\modif for the eclipsed star} and `tertiary' {\modif for
the eclipsing object}) were varied over a wide range and parametrized
by mass. For hierarchical triples the stellar components were
constrained to lie on the same isochrone, while for
background/foreground blends the properties of the intruding star were
taken from a representative solar-metallicity, 3\,Gyr isochrone.

The distance between the binary ($d_{\rm EB}$) and the target ($d_{\rm
\corotseven}$) was expressed for convenience in terms of the distance
modulus difference ($\Delta\delta = 5\log \left[d_{\rm EB}/d_{\rm
\corotseven}\right]$), {\modif which is independent of interstellar
extinction,} and varied between $-5$ and +10. The impact parameter was
allowed to vary between 0 and unity, and the mass of the secondary
stars ranged from 0.1\,$M_{\sun}$ to 1.4\,$M_{\sun}$. For stellar
tertiaries we explored masses from 0.1\,$M_{\sun}$ up to the mass of
the secondary; for tertiaries that are planets (assumed to be dark) we
allowed their sizes to be up to 2.0\,$R_{\rm Jup}$. We restricted our
blend simulations to circular orbits for all eclipsing pairs, as no
stellar or planetary systems with periods as short as that of
\corotseven\,b are known to have eccentric orbits, nor are expected to
from theoretical arguments \citep[see, e.g.,][]{Mazeh:08}.  Synthetic
light curves were generated with a detailed eclipsing binary code,
including proximity effects (tidal and rotational distortions),
limb-darkening, gravity brightening, and contamination from third
light. Differential extinction was included for chance alignments.

Blends providing poor fits compared to the \corotseven\ photometry in
a $\chi^2$ sense were considered to be ruled out. This enables us to
place constraints on the kinds of objects composing the eclipsing pair
that yield viable blends (i.e., acceptable fits), including their size
or mass, as well as other properties of the blend such as the overall
brightness and color.  For further details and applications of
\blender\ to other transiting planet candidates, we refer the reader
to the work of \cite{Torres:11}, \cite{Lissauer:11}, and
\cite{Fressin:11}.

\subsection{Background or foreground stars transited by a planet}

\begin{figure}
\epsscale{0.7}
\plotone{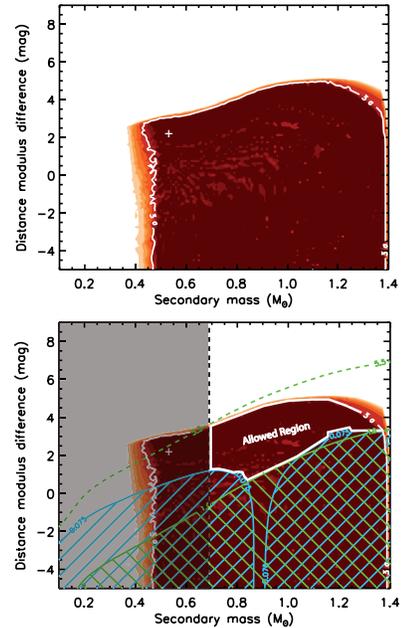}
\vskip 20pt
\begin{center}
\caption{ {\modif \emph{Top}}: Map of the $\chi^2$ surface (goodness
of fit) for blends involving background or foreground stars transited
by a larger planet. The vertical axis represents the distance between
the background pair of objects and the primary star, cast for
convenience in terms of the difference in the distance modulus
$\Delta\delta$ (note that $\Delta\delta$ is not equivalent to the
magnitude difference between the background system and the main star
because of the effects of differential extinction, which are included
in our simulations).  Only blends inside the solid white contour match
the \corot\ light curve within acceptable limits \citep[3$\sigma$,
where $\sigma$ is the significance level of the $\chi^2$ difference;
see][]{Fressin:11}.  {\modif Lighter-colored areas (red, orange,
yellow)} mark regions of parameter space giving increasingly worse
fits to the data (4$\sigma$, 5$\sigma$, etc.), and correspond to
blends we consider to be ruled out.
{\modif \emph{Bottom}: Same diagram as above with the addition of
observational constraints from follow-up measurements.  The constraint
from our \spitzer\ observations is represented by the shaded gray area
to the left of 0.69\,$M_{\sun}$; all false positives with secondary
masses smaller than 0.69\,$M_{\sun}$ can be rejected, as they lead to
transit depths at 4.5\,\micron\ that are inconsistent with our
measurements. Blends in the hatched green area are also ruled out
because they are bright enough to be detected spectroscopically
($\Delta m \leq 1.0$~mag, represented by the solid green line). The
hatched blue regions correspond to blends that can be excluded as well
because of their $r-K_s$ colors, which are either too red (left) or
too blue (right) compared to the measured value for \corotseven\ by
more than 3$\sigma$ (0.075~mag).  The combination of all of these
constraints leaves only a reduced area of parameter space (labeled
`Allowed Region') where blend models give tolerable fits to the
\corot\ light curve, and are not ruled out by any of the follow-up
observations. These blends are all brighter than $\Delta m = 5.5$~mag
(dashed green line). The white cross to the left of the Allowed Region
marks the location of a representative blend ruled out by \spitzer,
for which the predicted light curves are shown in Fig.~\ref{fig:blend}
(see text).
\label{fig:bp}}}
\end{center}
\end{figure}

We consider first the case of blends involving a background or
foreground star falling within the \corot\ aperture, and transited by
a larger planet. {\modif Because of the very short 0.8535-day orbital
period of \corotseven\ evolved stars (giants or subgiants) are ruled
out, so we focus in the following on main sequence stars.}  We find
that there is a very large range of spectral types (masses) permitted
for the secondary star, shown in {\modif top panel of}
Figure~\ref{fig:bp}, as well as a wide range of relative distances
between the eclipsing pair and the target. This is indicated by the
darker areas in the figure, delimited by the white contour
representing light curve fits that differ from the best transiting
planet model by a $\chi^2$ difference corresponding to a 3-$\sigma$
confidence level.  However, other constraints available for
\corotseven\ strongly restrict the number of these false
positives. {\modif This is illustrated in the bottom panel of the
figure, in which these additional constraints are superposed on the
same blend landscape displayed in the top panel.}  In particular, by
comparing the predicted $r-K_s$ color of each blend against the
measured color of the star from Exodat \citep[$r-K_s = 1.723 \pm
0.025$;][]{lege:09}, we find that a large fraction of the secondaries
with significantly different masses than the primary are ruled out
because the blends would be too red or too blue compared to the known
color index of \corotseven\ (by more than 3$\sigma$). {\modif These
excluded regions} are indicated by the blue hatched areas in {\modif
the bottom panel of} Figure~\ref{fig:bp}.  Other blends are excluded
because the secondary star would be very bright (within one magnitude
of the target), and would have been noticed spectroscopically, if
unresolved in the high-resolution imaging described earlier. {\modif
The section of parameter space excluded by this brightness criterion}
is indicated by the green hatched area, {\modif limited from above by}
the green diagonal line corresponding to a magnitude difference of
1.0~mag.  Note that, as mentioned in Sect.~\ref{sec:photometry}, this
is a very conservative brightness limit as a large fraction of stars
within 2 or even 3 magnitudes of \corotseven\ would most likely have
been detected spectroscopically. As indicated before, our \spitzer\
observations exclude secondary stars less massive than approximately
0.69\,$M_{\sun}$, which are common constituents of background blends.
This {\modif additional constraint} is shown by the shaded gray area
{\modif leftward of 0.69\,$M_{\sun}$. There is some overlap between
the gray region and the blue and green hatched areas, indicating that
those blends are excluded by more than one observational constraint.}

\begin{figure}
\epsscale{0.9}
\plotone{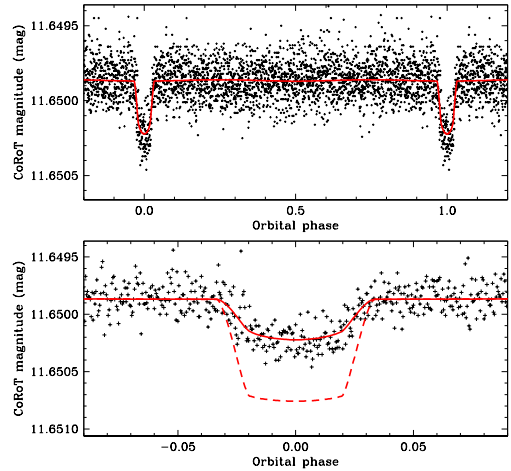}
\caption{ {\modif Example of a blend scenario (solid curves) that
reproduces the visible light curve from \corot\ (black crosses), but
that is ruled out solely by our \spitzer\ observations (and not by any
of the other follow-up observations). It involves an M0\,V star of
$0.53\,M_{\sun}$ with $V = 17.0$ in the background of the target star,
eclipsed by a Jovian planet with a radius 1.4 times that of
Jupiter. \emph{Top}: Folded light curve, full phase.  \emph{Bottom:}
Enlargement around the transit of \corotseven\,b.  The dashed curve
shows that the transit depth predicted for this same blend scenario in
the near-infrared 4.5\,\micron\ bandpass is much greater than in the
optical, which is inconsistent with our findings (see also
Sect.~\ref{sec:short_norm} and Fig.~\ref{folded}).}
\label{fig:blend}}
\end{figure}

An example of a blend that provides an acceptable fit to the \corot\
photometry is shown in Figure~\ref{fig:blend} (top panel).  This
scenario (location marked with a cross in Figure~\ref{fig:bp}) is not
excluded either by its color or its brightness, but \emph{is} clearly
ruled out by our \spitzer\ observations because it produces a much
deeper transit in the near-infrared (see bottom panel of
Figure~\ref{fig:blend}), which would have been easily detected. To
summarize, the combination of the \spitzer, color, and brightness
constraints removes many but not all blends involving a background or
foreground star transited by a larger planet. Those that remain reside
in the area of Figure~\ref{fig:bp} {\modif (bottom)} labeled ``Allowed
Region''.

\subsection{Background eclipsing binaries}

\begin{figure}
\vskip 10pt
\begin{center}
\epsscale{0.9}
\plotone{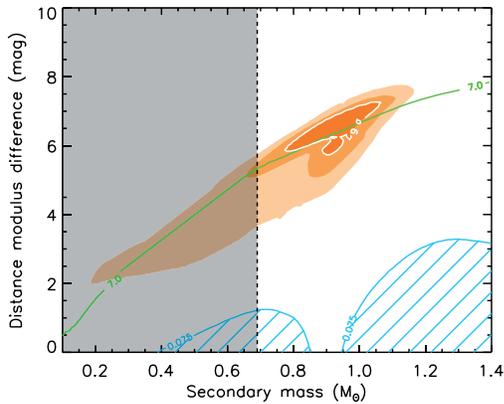}
\caption{Similar to Fig.~\ref{fig:bp} (and with the same color scheme)
for blends involving background or foreground stars transited by
another star.  Using \blender\ we find that no such background
eclipsing binary is able to reproduce the observed photometry at a
level better than 8.3\,$\sigma$ {\modif (white contour)} in comparison
to the $\chi^2$ of the best planet fit. Therefore, this excludes
\emph{all} background or foreground eclipsing binaries consisting of
two stars as potential false positives.
\label{fig:bs}}
\end{center}
\end{figure}

For the case of a background eclipsing binary composed of two stars,
interestingly we find that no combination of relative distance and
stellar properties for the eclipsing pair gives an acceptable fit to
the \corot\ light curve. The reason for this is that all such blend
configurations that can potentially reproduce the detailed shape of
the transit also lead to out-of-eclipse brightness changes
(ellipsoidal variations) with an amplitude that is not seen in the
data, and that are a consequence of the very short orbital period.
This implies that background blends of this kind can be confidently
ruled out, an important conclusion that does not follow from the
original \cite{lege:09} analysis.  Figure~\ref{fig:bs} shows that the
blends that yield the best fits to the photometry are excluded at the
8.3\,$\sigma$ level or higher. This result is significant, as it
substantially reduces the overall likelihood of blends for
\corotseven\,b.

We note that white dwarfs are also excluded as potential tertiaries.
Although the range of their radii overlaps with those of the planets
considered earlier in this section, their considerably larger mass
would once again induce significant ellipsoidal variation in the light
curve, which is ruled out by the observations. As was the case in the
previous section, giant stars do not constitute viable secondaries
because of the very short 0.8535-day period of the signal.

\subsection{Hierarchical triples}

\begin{figure}
\vskip 10pt
\begin{center}
\epsscale{0.9}
\plotone{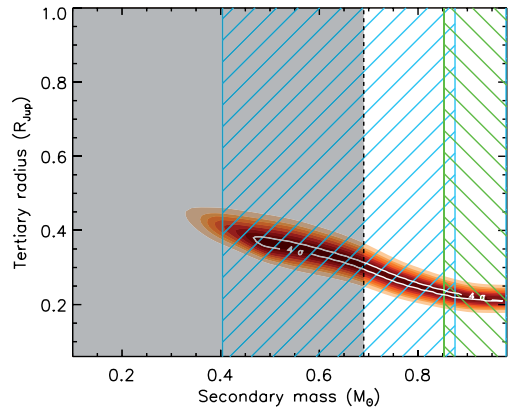}
\caption{Similar to Fig.~\ref{fig:bp} for the case of hierarchical
triple systems in which the secondary is transited by a planet. After
taking into account the \spitzer\ constraint ({\modif gray} shaded
region on the left), as well as the constraints on the $r-K_s$ color
and those on the brightness of the secondaries from spectroscopy (blue
and green hatched regions, respectively), we find that \emph{all}
triple configurations are excluded as false positives.\label{fig:htp}}
\end{center}
\end{figure}

Finally, for eclipsing binaries (consisting of two stars) that are
physically associated with the target in a hierarchical triple
configuration, we find that the blend light curves invariably have the
wrong shape to mimic a true transiting planet signal, for any
combination of stellar parameters for the secondary and tertiary.
Either the depth, duration, or the steepness of the ingress/egress
phases of the transits provide a very poor match to the \corot\
photometry, resulting in $\chi^2$ differences compared to the best
transit model corresponding to several hundred $\sigma$. These
scenarios are therefore all excluded.  On the other hand, if we allow
the tertiaries to be planets, the blend fits are somewhat better over
a wide range of masses for the secondaries when transited by a planet
of the appropriate size, but not quite at the 3$\sigma$ level or
better.  The best $\chi^2$ for a blend of this type corresponds to
about a 3.4$\sigma$ departure from a planet model. Although our formal
3-$\sigma$ limit is reasonable and consistent with common practice, it
is still somewhat arbitrary and one may argue that fits that are only
marginally worse might still be tolerable.  Even accepting this
possibility, Figure~\ref{fig:htp} shows that \emph{all} of these
blends are excluded by a combination of constraints from \spitzer,
color index, and brightness, even those that are 10$\sigma$ or more
away from the quality of a planet model.

\subsection{Summary of blends}

From the simulations described above, the only viable blend scenarios
for \corotseven\,b are those involving a larger planet transiting a
background or foreground star. \blender\ restricts those blends to the
area labeled ``Allowed Region'' in Figure~\ref{fig:bp}. These
configurations involve stars between 1.0 and 5.5~mag fainter than the
target within the \corot\ aperture.

\section{Validating \corotseven\,b}\label{sec:validation}

The {\it a priori\/} frequency of stars in the background or
foreground of the target that are orbited by a transiting planet and
are capable of mimicking the photometric signal may be estimated from
the density of stars in the vicinity of \corotseven, and the frequency
of transiting planets with the appropriate characteristics.  The
relevant area around the target is that in which stars of each
brightness would go undetected in the high-resolution imaging reported
by the \corot\ team. To obtain the number density (stars per square
degree) we make use of the Galactic structure models of
\cite{robi:03}, and we perform this calculation in half-magnitude
bins, as illustrated in Table~\ref{tab:blendfreq}. For each bin we
further restrict the star counts using the constraints on the mass of
the secondaries supplied by \blender\ (see Figure~\ref{fig:bp}).
These mass ranges are listed in column~3, and the resulting densities
appear in column~4.\footnote{\modif We note that the precipitous drop
in the numbers listed in column~4 below a magnitude of 16.7 is not due
to a real decrease in stellar density (which in fact rises for fainter
stars), but to the fact that the mass range allowed for blends in this
magnitude bin is significantly reduced.}  Bins with no entries
correspond to brightness ranges excluded by \blender.  For the maximum
angular separation ($\rho_{\rm max}$, column~5) at which stars of each
brightness would escape detection we have adopted the value 0\farcs4,
based on the report by \cite{lege:09} that companions outside this
range would have been seen in their VLT/NACO image down to a magnitude
difference of 6.5~mag compared to the target. We note that this
$\rho_{\rm max}$ value is a conservative limit, as stars at closer
separations would have been detected if they had smaller magnitude
differences, which would reduce the blend frequency.  The result for
the number of background stars in each magnitude bin is given in
column~6, in units of $10^{-6}$.

To estimate the frequency of transiting planets that might be expected
to orbit these stars (and lead to a false positive) we rely on the
results from \cite{boru:11}, who reported a total of 1235 planet
candidates among the 156,453 \kepler\ targets observed during the
first four months of the Mission. These signals have not yet been
confirmed to be caused by planets, and therefore remain candidates
until they can be thoroughly followed up. However, the rate of false
positives in this sample is expected to be quite small \citep[$10\%$
or less; see][]{mort:11}, so our results will not be significantly
affected by the assumption that all of the candidates are planets. We
further assume that the census of \cite{boru:11} is largely complete.
After accounting for the additional \blender\ constraint on the range
of planet sizes for blends of this kind (tertiaries of
0.24--1.42\,$R_{\rm Jup}$), we find that the transiting planet
frequency is $f_{\rm planet} = 571/156,\!453 = 0.0036$. Multiplying
this frequency by the star counts in column~6 of
Table~\ref{tab:blendfreq}, we arrive at a total blend frequency (BF)
listed at the bottom of column~7, ${\rm BF} = 4.2 \times 10^{-7}$.

This figure represents the {\it a priori\/} likelihood of a false
positive, and we note that it is approximately 3 orders of magnitude
smaller than indicated by the calculations of \cite{lege:09}. However,
we do not consider this to represent the ``false alarm rate'', as the
expected likelihood of a transiting planet of the characteristics
implied by the transit signal is also very small.  We adopt here a
Bayesian approach analogous to that employed to validate previous
\kepler\ candidates, in which our confidence in the planetary nature
of the signal will depend on how the blend likelihood compares to the
{\it a priori\/} likelihood of a true transiting planet (PF),
addressed below. {\modif Thus, we seek to estimate the \emph{odds
ratio} PF/BF.}  This is a significant conceptual difference compared
to the frequentist approach by \cite{lege:09}, who considered only the
likelihood of a blend (BF).

{\modif Implicit in the blend frequency calculation above (${\rm BF} =
  4.2 \times 10^{-7}$) is the 3$\sigma$ criterion on the quality of
  the light curve fit relative to a transit model fit that we used as
  a condition for a blend scenario to be acceptable (see
  Sect.~\ref{sec:blender}). For a fair comparison, we use a similar
  3$\sigma$ criterion to establish the {\it a priori\/} transiting
  planet frequency (PF), for the numerator of our odds ratio. We
  estimate PF} by counting the \kepler\ candidates in the
\cite{boru:11} sample that have radii within 3$\sigma$ of the value
determined from the best fit to the \corotseven\ data \citep[$R_p =
  1.58 \pm 0.10\,R_{\earth}$;][]{brun:10}.  We find 231 candidates
within this range, giving a planet frequency ${\rm PF} = 231/156,\!453
= {\modif 0.0015}$.

Thus, the likelihood of a planet is more than 3,500 times greater than
that of a false positive (${\rm PF/BF} = 0.0015/4.2\times10^{-7}$),
which we consider sufficient to independently validate \corotseven\,b
as a true planet with a high degree of confidence. We note that our
blend frequency calculation assumes the 1,235 candidates cataloged by
\cite{boru:11} are all true planets.  If we were to assume
conservatively that as many as 50\% are false positives in the radius
range specific to the transiting planet case \citep[an unlikely
proposition that is also inconsistent with other evidence;
see][]{howa:10, boru:11}, the planet likelihood would still be nearly
1,800 times greater than the likelihood of a blend, implying a false
alarm rate sufficiently small to validate the signal.

Additionally, in estimating the {\it a priori} true planet frequency
{\modif (PF)} and the frequency of larger planets involved in blends,
we have not {\modif placed any restriction on the periods of either
kind of planet. One may expect, for example, that limiting the periods
of larger planets involved in blends to be similar to that of
\corotseven\,b, which is quite short, might reduce the blend frequency
quite considerably. On the other hand, a similar period limit imposed
on the true planet frequency would decrease PF as well, though these
effects may not be equal. Thus, it is possible that the odds ratio
would be altered. As it turns out, our calculations without any period
constraints are the most conservative.}  For example, limiting the
periods to be shorter than 5 days, the census of \cite{boru:11}
indicates that larger planets are comparatively less common than the
smaller ones, resulting in an improvement in the PF/BF odds ratio of a
factor of 2.2.  Restricting the periods to be less than 3 days, the
odds ratio improves by a factor of 2.8.

\section{Discussion}

Our observation of four transits of \corotseven\,b with the \spitzer\
Space Telescope at a wavelength of 4.5\,\micron\ has resulted in the
detection of the transit with a depth of $0.426 \pm 0.115$\,mmag,
which is consistent with the \corotseven\,b planet scenario described
in \cite{lege:09}. Although the signal-to-noise ratio of this infrared
detection is relatively low, it has been obtained with a now standard
treatment of the \spitzer\ data with no {\it a priori\/} knowledge of
the transit parameters (aside from the timing windows in order to plan
the four \spitzer\ visits).

Warm \spitzer\ is currently the only facility available that has the
capability of detecting such shallow transits at wavelengths that are
sufficiently separated from the \corot\ (or \kepler) passbands to be
helpful. In this case the observations were successful, and the
transit at $4.5~\micron$ is shown to have virtually the same depth as
in the optical. This places a strong constraint on the color of
potential blends, which are restricted to have secondaries of similar
spectral type as the primary star. In the case of \corotseven\,b this
allows us to rule out most cool stars (which constitute the majority
of background stars) as potential contaminants.

The detailed analysis of the \corotseven\ photometry with \blender\
combined with constraints from other observations eliminates the vast
majority of possible blend scenarios.  This includes all background
eclipsing binaries composed of two stars, most of the scenarios
involving chance alignments with a star transited by a larger planet,
and all hierarchical triple configurations. The remaining scenarios
are much less likely (by a factor of 3,500) than a true transiting
planet, thereby validating the planetary nature of the signal with
very high confidence. We point out that this conclusion has been
reached with very conservative assumptions regarding some of the
observational constraints. For example, we have ignored the fact that
the high-resolution VLT/NACO imaging by \cite{lege:09} permits the
detection of background stars closer than our adopted limit of
$\rho_{\rm max} = 0\farcs4$ from the target if they are relatively
bright. Additional imaging by those authors with FASTCAM on the 2.5\,m
NOT telescope provides even tighter constraints that we have not used,
reaching down to sensitivities of 4 magnitudes fainter than the target
at 0\farcs18.  We have also not considered the full potential of
spectroscopy to rule out closer companions; we have assumed only that
stars within 1.0 magnitude of the target would have been identified,
whereas in reality the sensitivity of those observations (not only the
near-infrared CRIRES spectrum, but also those from HARPS) is probably
much greater ($\Delta m \approx$ 2--3\,mag). Furthermore, we have not
made use of other information reported by \cite{lege:09}, such as the
constraints from the red, green, and blue passbands into which the
light from the \corot\ instrument can be split, which can further
limit the pool of potential false positives. Incorporating all of
these constraints can only reduce the blend frequency, resulting in an
even greater confidence level for the validation of \corotseven\,b as
the first super-Earth.

Our \spitzer\ observations have also significantly improved our
knowledge of the ephemeris of the first transiting super-Earth.
Transits of \corotseven\,b have not been observed with any other
photometric facility since the end of the LRa01 field observations by
the \corot\ satellite. Given the formal uncertainty in the orbital
period reported by \cite{lege:09}, the accumulated error in the
predicted times of transit up to the date of our \spitzer\
observations is approximately 20\,min, or about a quarter of the
transit duration.  Here we have improved the period determination by
about a factor of four, to $P = 0.853590 \pm 0.000006$ days, ensuring
the transits will be recoverable for at least another decade (provided
there are no physical mechanisms operating to change the period).

The \kepler\ satellite has recently released a large number of very
promising shallow transit candidates \citep{boru:11}. From the high
quality of this photometry the detailed transit shapes are often very
well defined, increasing the power of tools such as \blender\ that
make use of that information to rule out blend scenarios that result
in poor fits to the data. The exquisite relative astrometric precision
delivered by the \kepler\ instrument provides important additional
constraints from an analysis of the motion of the photocenters of the
images in and out of transit \citep[see, e.g.,][]{Batalha:10}, which
can typically exclude 90\% or more of unresolved stars falling within
the photometric aperture that can potentially contaminate the flux
measurements. Spectroscopy and high-resolution imaging contribute
further valuable constraints. These tools have already been used to
validate small signals for which radial-velocity confirmation is
currently out of reach, including possibly rocky planets. Examples
include the super-Earth Kepler-9\,d \citep{Torres:11}, and the
Neptune-size planets Kepler-10\,c \citep{Fressin:11}, Kepler-11\,g
\citep{Lissauer:11}, {\modif and Kepler-19\,b \citep{Ballard:11}}.

\spitzer\ observations provide a very effective way of ruling out a
large fraction of the blend scenarios involving a background star with
a different effective temperature (or color) than the target.  These
chance alignments are typically the most serious concern regarding the
nature of the transit signals. Among the most interesting candidates
expected from the \kepler\ Mission are rocky planets in the habitable
zone of their parent stars. The typically longer periods of these
objects mean that methods of confirmation relying on the dynamical
influence of the planet and/or the quality of the phase-folded
photometry (obtained by summing individual transit events) will be
more problematic, as their efficiency scales down with orbital period.
We anticipate that the \spitzer\ telescope in its warm phase will
prove critical for validating such objects, as its efficiency does not
depend on period, but relies instead on a different intrinsic property
of the transits, which is their achromaticity.

\acknowledgements

We are grateful to the anonymous referee for many very helpful
comments and suggestions.  This work is based on observations made
with the \spitzer\ Space Telescope, which is operated by the Jet
Propulsion Laboratory, California Institute of Technology, under
contract to NASA.  Support for this work was provided by NASA through
an award issued by JPL/Caltech. This research has made use of the
facilities at the NASA Advanced Supercomputing Division (NASA Ames
Research Center).




\begin{deluxetable}{ccccccc}
\tabletypesize{\scriptsize} 
\tablewidth{0pc} 

\tablecaption{Blend frequency estimate for \corotseven\,b. \label{tab:blendfreq}}
\tablehead{ 
& & 
\multicolumn{5}{c}{Blends Involving Planetary Tertiaries} \\[+0.5ex] 
\cline{3-7} \\ [-1.5ex]
\colhead{Magnitude Range} &
\colhead{$\Delta m$} &
\colhead{Stellar} &
\colhead{Stellar Density} &
\colhead{$\rho_{\rm max}$} &
\colhead{Stars} &
\colhead{Transiting Planets} \\
\colhead{(mag)} &
\colhead{(mag)} &
\colhead{Mass Range} &
\colhead{(per sq.\ deg)} &
\colhead{(\arcsec)} &
\colhead{($\times 10^{-6}$)} &
\colhead{0.24--1.42\,$R_{\rm Jup}$, $f_{\rm planet}=0.36$\%} \\
\colhead{} &
\colhead{} &
\colhead{ {\modif  ($M_{\odot}$) }  } &
\colhead{} &
\colhead{} &
\colhead{} &
\colhead{($\times 10^{-6}$)} \\
\colhead{(1)} &
\colhead{(2)} &
\colhead{(3)} &
\colhead{(4)} &
\colhead{(5)} &
\colhead{(6)} &
\colhead{(7)}
}
\startdata
11.7--12.2  &  0.5 & \nodata   & \nodata&\nodata & \nodata& \nodata \\
12.2--12.7  &  1.0 & \nodata   & \nodata&\nodata & \nodata& \nodata \\
12.7--13.2  &  1.5 & 0.84--1.36  & 91    &  0.40  &  3.53  & 0.013 \\
13.2--13.7  &  2.0 & 0.82--1.33  & 137   &  0.40  &  5.31  & 0.019 \\
13.7--14.2  &  2.5 & 0.77--1.30  & 207   &  0.40  &  8.03  & 0.029 \\
14.2--14.7  &  3.0 & 0.69--1.26  & 263   &  0.40  &  10.2  & 0.037 \\
14.7--15.2  &  3.5 & 0.69--1.22  & 383   &  0.40  &  14.9  & 0.054 \\
15.2--15.7  &  4.0 & 0.69--1.17  & 515   &  0.40  &  20.0  & 0.072 \\
15.7--16.2  &  4.5 & 0.69--1.08  & 569   &  0.40  &  22.0  & 0.081 \\
16.2--16.7  &  5.0 & 0.69--0.99  & 637   &  0.40  &  24.7  & 0.090 \\
16.7--17.2  &  5.5 & 0.69--0.80  & 186   &  0.40  &  7.21  & 0.026 \\
17.2--17.7  &  6.0 & \nodata     & \nodata&\nodata & \nodata& \nodata \\
17.7--18.2  &  6.5 & \nodata     & \nodata&\nodata & \nodata& \nodata \\
18.2--18.7  &  7.0 & \nodata     & \nodata&\nodata & \nodata& \nodata \\
18.7--19.2  &  7.5 & \nodata     & \nodata&\nodata & \nodata& \nodata \\
19.2--19.7  &  8.0 & \nodata     & \nodata&\nodata & \nodata& \nodata \\
19.7--20.2  &  8.5 & \nodata     & \nodata&\nodata & \nodata& \nodata \\
\noalign{\vskip 6pt}
\multicolumn{2}{c}{Totals} & & 2988 & & 115.9  & 0.42 \\
\noalign{\vskip 4pt}
\hline
\noalign{\vskip 4pt}
\multicolumn{7}{c}{Total frequency (BF) = $4.2 \times 10^{-7}$} \\ 
\enddata

\tablecomments{Magnitude bins with no entries correspond to
brightness ranges in which \blender\ excludes all blends.}

\end{deluxetable}
\vskip 35pt


\begin{thebibliography}{}

\bibitem[Aigrain \& Irwin(2004)]{aigr:04} Aigrain, S., \& Irwin, M.\ 2004, \mnras, 350, 331
\bibitem[Baglin et al.(2002)]{bagl:02} Baglin, A., Auvergne, M., Barge, P., Buey, J.-T., Catala, C., Michel, E., Weiss, W., \& COROT Team 2002, Stellar Structure and Habitable Planet Finding, 485, 17
\bibitem[Ballard et al.(2011)]{Ballard:11} Ballard, S.\ et al.\ 2011, \apj, in press (arXiv:1109.1561)
\bibitem[Batalha et al.(2010)]{Batalha:10} Batalha, N.\ M.\ et al.\ 2010, \apj, 713, L103
\bibitem[Batalha et al.(2011)]{bata:11}  Batalha, N.\ M.\ et al.\ 2011, \apj, 729, 27
\bibitem[Berta et al.(2011)]{bert:11} Berta, Z.\ K., 
Charbonneau, D., Bean, J., Irwin, J., Burke, C.\ J., D{\'e}sert, J.-M., 
Nutzman, P., \& Falco, E.\ E.\ 2011, \apj, 736, 12 
\bibitem[Boisse et al.(2011)]{Boisse:11} Boisse, I., Bouchy, F., H\'ebrard, G., Bonfils, X., Santos, N., \& Vauclair, S. 2011, \aap, 528, 4
\bibitem[Borucki et al.(2011)]{boru:11}  Borucki, W.\ J.\ et al.\ 2011b, arXiv:1102.0541 
\bibitem[Bruntt et al.(2010)]{brun:10} Bruntt, H.\ et al.\ 2010, \aap, 519, A51  
\bibitem[Charbonneau et al.(2005)]{char:05} Charbonneau, D.\ et al.\ 2005, \apj, 626, 523
\bibitem[Charbonneau et al.(2008)]{char:08} Charbonneau, D.\ et al.\ 2008, \apj, 686, 1341
\bibitem[Charbonneau et al.(2009)]{char:09} Charbonneau, D.\ et al.\ 2009, \nat, 462, 891 
\bibitem[Claret(2000)]{Claret:00} Claret, A. 2000, \aap, 363, 1081
\bibitem[Deming et al.(2005)]{demi:05} Deming, D., Seager, S., Richardson, L.\ J., \& Harrington, J.\ 2005, \nat, 434, 740 
\bibitem[Fazio et al.(2004)]{fazi:04} Fazio, G.\ G.\ et al., 2004, \apjs, 154, 10
\bibitem[Ferraz-Mello et al.(2011)]{Ferraz-Mello:11} Ferraz-Mello, S., Tadeu dos Santos, M., Beaug\'e, C., Michtchenko, T.\ A., \& Rodr\'i\i guez, A. 2011, \aap, 531, 161
\bibitem[Ford(2005)]{ford:05} Ford, E.\ B.\ 2005, \aj, 129, 1706
\bibitem[Fressin et al.(2009)]{fres:09} Fressin, F., Guillot, T., \& Nesta, L.\ 2009, \aap, 504, 605
\bibitem[Fressin et al.(2011)]{Fressin:11} Fressin, F.\ et al.\ 2011, \apj, in press 
\bibitem[Girardi et al.(2000)]{gira:00} Girardi, L., Bressan, A., Bertelli, G., \& Chiosi, C.\ 2000, \aaps, 141, 371
\bibitem[Hatzes et al.(2010)]{hatz:10} Hatzes, A.\ P.\ et al.\ 2010, \aap, 520, A93
\bibitem[Hatzes et al.(2011)]{hatz:11} Hatzes, A.\ P.\ et al.\ 2011, arXiv:1105.3372 
\bibitem[Howard et al.(2010)]{howa:10} Howard, A.\ W.\ et al.\ 2010, Science, 330, 653  
\bibitem[Howell et al.(2011)]{howe:11} Howell, S.\ B.\ et al.\ 2011, submitted to AJ. 
\bibitem[Jenkins et al.(2010)]{jenk:10} Jenkins, J.\ M.\ et al.\ 2010, arXiv:1001.0416
\bibitem[Knutson et al.(2008)]{knut:08} Knutson, H.\ A., Charbonneau, D., Allen, L.\ E., Burrows, A., \& Megeath, S.\ T. 2008, \apj, 673, 526
\bibitem[Knutson et al.(2009)]{knut:09} Knutson, H.\ A., Charbonneau, D., Burrows, A., et al.\ 2009, \apj, 655, 566
\bibitem[L{\'e}ger et al.(2009)]{lege:09} L{\'e}ger, A.\ et al.\ 2009, \aap, 506, 287 
\bibitem[Lissauer et al.(2011)]{Lissauer:11} Lissauer, J.\ J.\ et al.\ 2011, \nat, 470, 53
\bibitem[Markwardt(2009)]{mark:09} Markwardt, C.\ B.\ 2009, Astronomical Society of the Pacific Conference Series, 411, 251 
\bibitem[Mandel \& Agol(2002)]{mand:02} Mandel, K. \& Agol, E. 2002, \apj, 580, L171
\bibitem[Mazeh(2008)]{Mazeh:08} Mazeh, T. 2008, in Tidal Effects in Stars, Planets and Disks, EAS Publications Series, eds.\ M.-J.\ Goupil \& J.-P.\ Zahn (EDP Sciences), Vol.\ 29, p. 1
\bibitem[Morton \& Johnson(2011)]{mort:11} Morton, T.\ D., \& Johnson, J.\ A. 2011, arXiv:1101.5630 
\bibitem[Morales-Calderon et al.(2006)]{mor:06} Morales-Calderon, M.\ et al.\ 2006, \apj, 653, 1454
\bibitem[Pont et al.(2006)]{pont:06} Pont, F., Zucker, S., \& Queloz, D. 2006, \mnras, 373, 231 
\bibitem[Pont et al.(2011)]{pont:11} Pont, F., Aigrain, S., \& Zucker, S. 2011, \mnras, 411, 1953
\bibitem[Queloz et al.(2009)]{quel:09} Queloz, D.\ et al.\ 2009, \aap, 506, 303 
\bibitem[Reach et al.(2005)]{reac:05} Reach, W. T.\ et al.\ 2005, PASP, 117, 978
\bibitem[Robin et al.(2003)]{robi:03} Robin, A.\ C., Reyl\'e, C., Derri\'ere, S., \& Picaud, S. 2003, \aap, 409, 523
\bibitem[Sing(2010)]{Sing:10} Sing, D.\ K. 2010, \aap, 510, 21
\bibitem[Torres et al.(2004)]{Torres:04} Torres, G., Konacki, M., Sasselov, D.\ D., \& Jha, S.\ 2004, \apj, 614, 979 
\bibitem[Torres et al.(2011)]{Torres:11} Torres, G.\ et al.\ 2011, \apj, 727, 24 
\bibitem[Werner et al.(2004)]{wern:04} Werner, M.\ W. et al., 2004, \apjs, 154, 1
\bibitem[Winn et al.(2007)]{winn:07} Winn, J.\ N., Holman, M.\ J., \& Fuentes, C.\ I. 2007, \aj, 133, 11
\bibitem[Winn et al.(2009)]{winn:09} Winn, J.\ N., Johnson, J.\ A., Albrecht, S., Howard, A.\ W., Marcy, G.\ W., Crossfield, I.\ J., \& Holman, M.\ J. 2009, \apjl, 703, L99 

\end{thebibliography}
\end{document}